\documentclass[12pt]{article}

\usepackage{graphicx}
\usepackage{epsfig}
\begin{document}


\def\lsim{\mathrel{\raise3pt\hbox to 8pt{\raise -6pt\hbox{$\sim$}\hss {$<$}}}} 
\newcommand{\be}{\begin{equation}}
\newcommand{\ee}{\end{equation}}
\newcommand{\bea}{\begin{eqnarray}}
\newcommand{\eea}{\end{eqnarray}}
\newcommand{\da}{\dagger}
\newcommand{\dg}[1]{\mbox{${#1}^{\dagger}$}}
\newcommand{\hlf}{\mbox{$1\over2$}}
\newcommand{\lfrac}[2]{\mbox{${#1}\over{#2}$}}
\newcommand{\scsz}[1]{\mbox{\scriptsize ${#1}$}}
\newcommand{\tsz}[1]{\mbox{\tiny ${#1}$}}
\newcommand{\doref}{\bf (*** REF.??? ***)}



\begin{flushleft}

\Large{\bf Search for a Solution of the Pioneer Anomaly}

\vspace{0.5in}

\normalsize
\bigskip 

{\bf Michael Martin Nieto$^a$ and John D. Anderson$^b$}  \\

\normalsize
\vskip 15pt

${^a}$Theoretical Division (MS-B285), Los Alamos National Laboratory,\\
Los Alamos, New Mexico 87545, U.S.A.\footnote{ 
Email: {\tt mmn@lanl.gov}} \\
\vspace{0.25in}
${^b}$Global Aerospace Corporation, 711 W. Woodbury Road Suite H, \\
Altadena, Ca. 91001, U.S.A.\footnote{
Email:{\tt John.D.Anderson@gaerospace.com}}

\end{flushleft}


\begin{abstract}

In 1972 and 1973 the Pioneer 10 and 11 missions were launched.  They were the first to explore the outer solar system and achieved stunning breakthroughs in deep-space exploration.  But beginning in about 1980 an unmodeled force of $\sim 8 \times 10^{-8}$ cm/s$^2$, directed approximately towards the Sun,  appeared in the tracking data.  It later was unambiguously verified as being in the data and not an artifact. The cause remains unknown (although radiant heat remains a likely origin).  With time more and more effort has gone into understanding this anomaly (and also possibly related effects).  We review the situation and describe ongoing programs to resolve the  issue.  
\end{abstract}

PACS \\
95.10.Eg,    (Orbit determination and improvement), \\
95.55.Pe,   (Lunar, planetary, and deep-space probes),   \\
96.12.De    (Orbital and rotational dynamics)


\newpage


\section{The Pioneer Missions}
\label{intro}

In the 1960's the use of planetary flybys for gravity assists of spacecraft became of wide interest.  That is when the Jet Propulsion Laboratory (JPL) first started thinking about what became the ``Grand Tours" of the 1970's and 1980's (the Voyager missions).  The concept was to use flybys of the major planets to both modify the direction of a spacecraft and also to add to its heliocentric velocity.  

At the time many found it surprising that energy could be transferred to a spacecraft from the orbital-motion angular-momentum of a planet about the Sun. This was despite the fact it had been known since the works of Lagrange, Jacobi, and Tisserand on the three-body problem 
\cite{moulton, danby} that the energies of comets could be affected by passing near Jupiter.
Even in the simplest, circular restricted 3-body problem \cite{danby}, it is {\it not} that the energy of each object is conserved, only that the total energy of the entire system is conserved.  Flybys can both give kinetic energy to a spacecraft (to boost its orbital velocity) and also can take kinetic energy from it (to slow it down for an inner body encounter).

\begin{figure}[ht] 
    \noindent
    \begin{center}  
\includegraphics[width=2.0in]{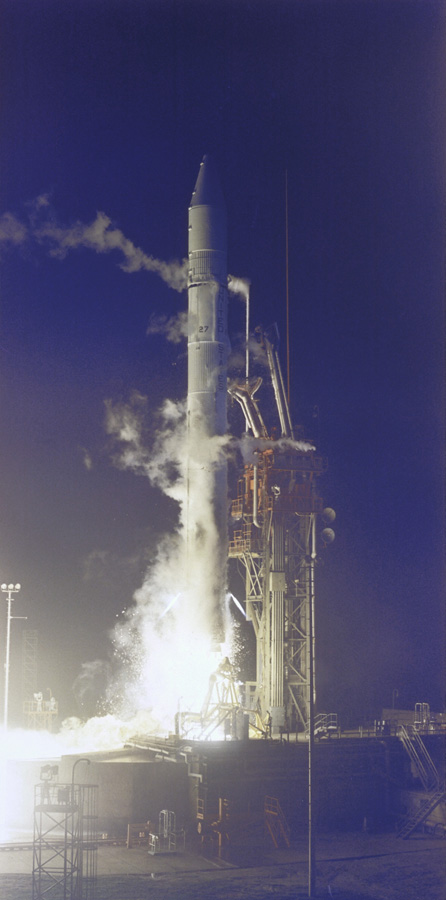}
\caption{\small{Pioneer 10's launch on 2 March 1972.   
}
 \label{pio10launch}}
    \end{center}
\end{figure}

The first missions to fly to deep space were the Pioneers.  By using flybys, heliocentric velocities were obtained that were unfeasible at the time by using only chemical fuels.
Pioneer 10 was launched on 2 March 1972 local time, aboard an Atlas/Centaur/TE364-4 launch vehicle 
(see Figure \ref{pio10launch}).
It was the first craft launched into deep space and was the first to reach an outer giant planet, Jupiter, on 4 Dec. 1973 
\cite{fimmel,wolverton}.  
Later it was the first spacecraft to leave the ``solar system" (past the orbit of Pluto, as it was then defined).  The Pioneer project eventually extended over decades.  It was managed at NASA/Ames Research Center under the hands of four successive project managers; the legendary Charlie Hall, Richard Fimmel, Fred Worth, and the current Larry Lasher. 


While in its Earth-Jupiter cruise, Pioneer 10 was still bound to the solar system.  By 9 January 1973 Pioneer 10 was at a distance of 3.40 AU (Astronomical Units\footnote{
An Astronomical Unit is the mean Sun-Earth distance, about 150,000,000 km.}), 
beyond the asteroid belt.  This in itself was a happy surprise.  The craft had not been destroyed passing through the asteroid belt, which had been greatly feared at the time.  

With the Jupiter flyby, Pioneer 10 reached escape velocity from the solar system.
Pioneer 10 has an asymptotic escape velocity from the Sun of 11.322 km/s (2.388 AU/yr).  It is  headed in the general direction of the star Aldebaran (scheduled to reach that region in about 2 million years).  This direction is opposite the relative motion of the solar system in the local interstellar dust cloud and opposite to the direction towards the galactic center.

Pioneer 11 followed soon after Pioneer10, with a launch on 6 April 1973.  It, too, cruised  to Jupiter on an approximate heliocentric ellipse.  This time during the Earth-Jupiter cruise, it was determined that a carefully executed flyby of Jupiter could put the craft on a trajectory to encounter Saturn in 1979.  So, on 2 Dec. 1974, when Pioneer 11 reached Jupiter, it underwent a Jupiter gravity assist that sent it back inside the solar system to catch up with Saturn on the far side.   It was then still on an ellipse, but a more energetic one.  Pioneer 11 reached as close to the Sun as 3.73 AU on 2 February 1976.  

Pioneer 11 reached Saturn on 1 Sept. 1979.  The trajectory took the craft under the ring plane on approach.  After passing through the plane -- again without catastrophic consequences -- Pioneer 11 came within 24,000 km of Saturn.  Then Pioneer 11 embarked on an escape hyperbolic trajectory with an asymptotic escape velocity from the Sun of 10.450 km/s ( 2.204 AU/yr).  

\begin{figure}[h!] 
    \noindent
    \begin{center}  
\includegraphics[width=3.75in]{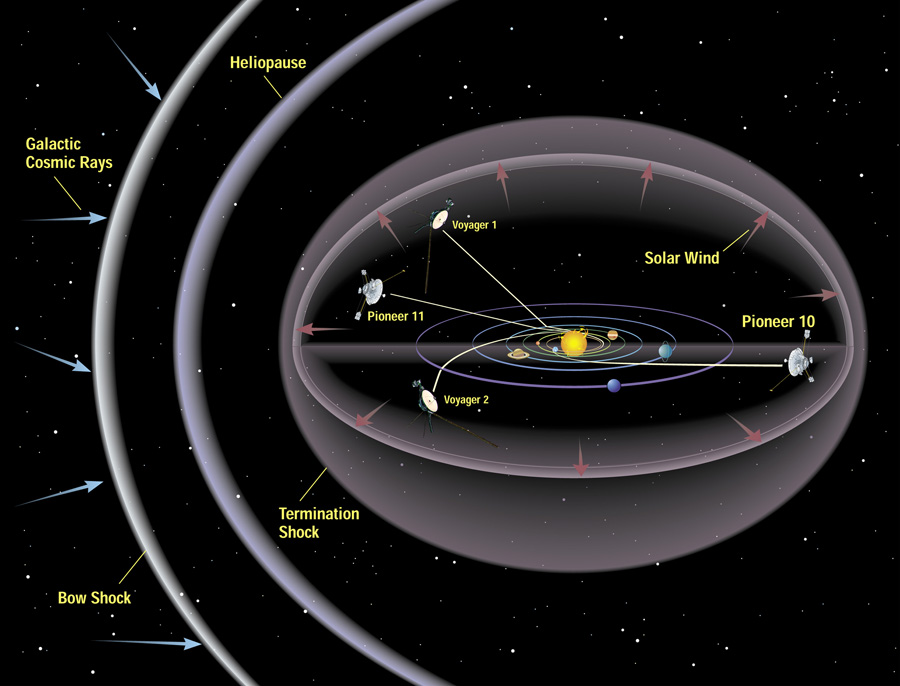}
\caption{\small{View of the Pioneer 10, Pioneer 11, and Voyager trajectories.  Pioneer 10 is traveling in a direction almost opposite to the galactic center, while Pioneer 11 is heading approximately in the closest direction to the heliopause.  The direction of the solar system's motion in the galaxy is approximately out of the drawing in the ecliptic.
[Digital artwork by T. Esposito. NASA ARC Image \# AC97-0036-3.]
}
\label{piopathdraw}}
\end{center}
\end{figure}

Pioneer 11 is headed in the direction of the Constellation Aquila, scheduled to arrive in about 4 million years.  This is approximately in the direction of the Sun's relative motion in the local interstellar dust cloud (towards the heliopause).  Therefore, its direction is roughly anti-parallel to the direction of Pioneer 10.
In Figure \ref{piopathdraw} 
the trajectories of the Pioneers ands Voyagers over the entire solar system are shown. 

\newpage

{\bf (*** LOCATION OF SIDE BAR I ***)}


\section*{SIDE BAR I:  Flybys}
\label{sideI}

{\it
In Flandro \cite{flandroref} and elsewhere \cite{jdaEGA, VA} the flyby process has been simply and intuitively described.  
Let the simple vector velocities in the heliocentric system be added to the orbital velocity of the planet (taken to be constant).  
(See Figure \ref{flandrofig}.) 


\begin{figure}[h] 
    \noindent
    \begin{center}  
\includegraphics[width=3.75in]{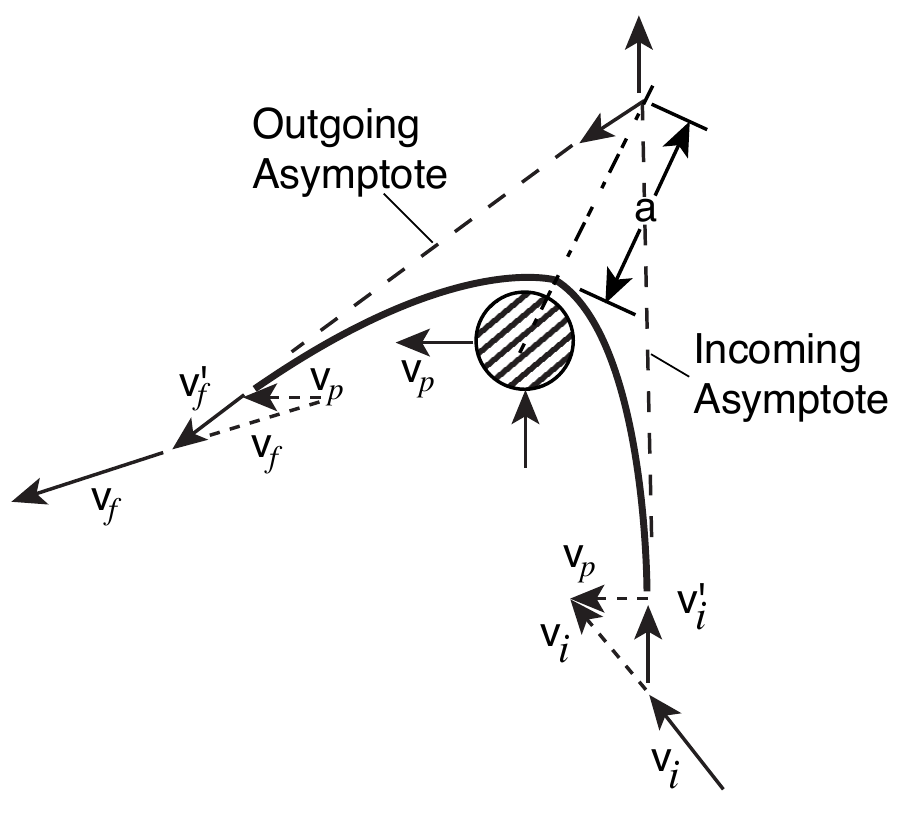}
\caption{Geometry of a flyby, modified from \cite{flandroref}. 
}
\label{flandrofig}
\end{center}
\end{figure}


\hspace{.25in} The initial and final velocities in the heliocentric system are ${\bf v}_i$ and ${\bf v}_f$.    The initial and final velocities in the planetary system are ${\bf v}'_i$ and ${\bf v}'_f$.  The velocity of the planet in the solar system is ${\bf v}_p$.  The change in kinetic energy per unit mass is 
\begin{equation}
\Delta {\cal K} = ({\bf v}_f \cdot {\bf v}_f - {\bf v}_i\cdot {\bf v}_i)/2.  \label{fly1}
\end{equation}
A little algebra \cite{flandroref,VA} gives one
\begin{equation}
\Delta {\cal K} = {\bf v}_p \cdot ({\bf v}_f' - {\bf v}_i').                 \label{fly2}
\end{equation}

\hspace{.25in}Roughly speaking, in the planetary system which rotates anticlockwise, if a satellite in the ecliptic comes from inside the planetary orbit, travels behind the planet, and then goes around it counter-clockwise, kinetic energy will be added to the spacecraft.  Contrarily if a satellite comes from inside the planetary orbit, travels in front of the planet, and then goes around it clockwise, kinetic energy will be taken away from the spacecraft. 

\hspace{.25in} Of course one is not violating conservation of energy.  The energy (and angular-momentum change) is absorbed by the planet that is  being flown by.  However, for such a massive body the relatively infinitesimal change is not noticeable.  Further, there {\it is} (in high approximation) a conserved quantity for the spacecraft in the barycentric system, Jacobi's integral \cite{moulton,danby}:
\be
J = - C/2 = ({\cal K} + {\cal V}) + {\cal L} 
          = ({\cal E}) - \mathbf{\omega \hat{z} \cdot r \times v}, \label{jacobi}
\ee
where $\{{\cal V},{\cal L},{\cal E}\}$ are the potential, rotational-potential, and total energies, respectively, per unit mass, $\mathbf{\omega}$ is the angular velocity of the planet (system) whose vector is aligned with $\mathbf{\hat{z}}$, the unit vector normal to planet's rotational plane.  
Eq. (\ref{jacobi}) is exactly a constant  in the circular restricted 3-body problem, and shows how kinetic energy can be exchanged with angular momentum during a flyby.  
}



\section{The Pioneer Navigation}
\label{navigation}

\begin{figure}[h!] 
    \noindent
    \begin{center}  
\includegraphics[width=3.75in]{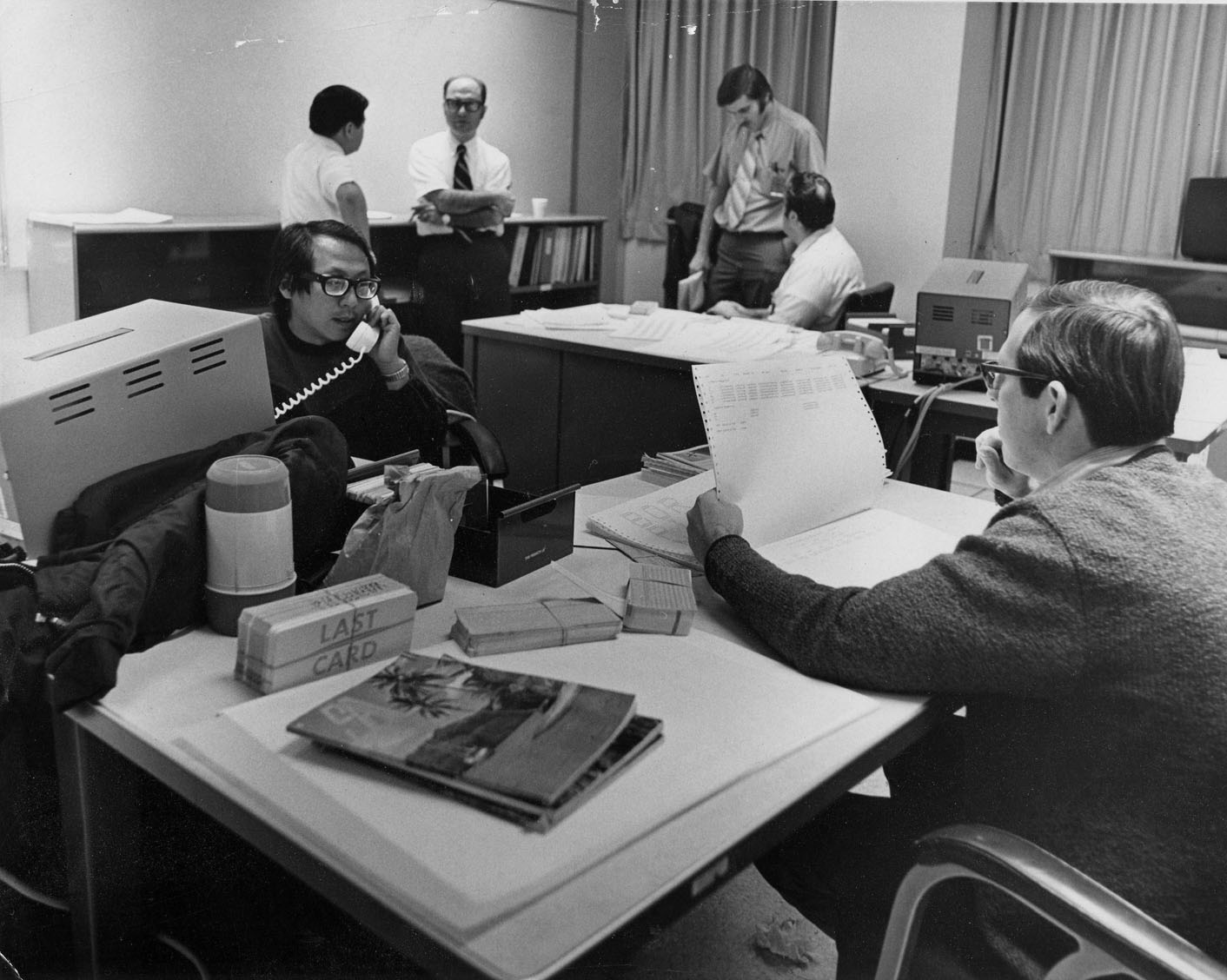}
\caption{Members of the JPL navigation team working on the day of Pioneer 10's launch.  In the foreground are Tony Liu and Phil Laing, who later became part of the Pioneer anomaly Collaboration.  In the background are Sun Kuen Wong, Jack Hohikian, Steve Reinbold, and Bruce O'Reilly.  Note the stack of computer program cards labeled “LAST CARD,” the large format computer printout paper, and the Tektronix scope, evidence of the technologies used then.  
\label{phil}}
\end{center}
\end{figure}

The Pioneer navigation was carried out at the Jet Propulsion Laboratory.  It was ground-breaking in its advances -- no craft had delved so far out into the solar system while aiming for such precise encounters.  To succeed the navigation team needed to correct imperfections in the existing codes in real time, as the craft was starting its journey  
(See Figure \ref{phil}.)   
Some of the crises overcome through personal interventions that could have cost employment remain part of the folk lore of this period.  But in the end, the team succeeded.  


They used NASA's Deep Space Network (DSN) to transmit and obtain the raw radiometric data.  An S-band signal ($\sim$2.11 Ghz) was sent up via a DSN antenna located either at Goldstone, California, outside Madrid, Spain, or outside Canberra, Australia.  On reaching the craft the signal was transponded back with a (240/221) frequency ratio ($\sim$2.29 Ghz), 
and received back at the same station (or at another station if, during the radio round trip, the original station had rotated out of view).  There the signal was compared with 240/221 times the recorded transmitted frequency and any Doppler frequency shift was measured directly by cycle count compared to an atomic clock. 
The processing of the raw cycle count produced a data record of Doppler frequency shift as a function of time, and from this a trajectory was calculated.  This procedure was done iteratively for purposes of converging to a best fit by nonlinear weighted least squares (minimization of the chi squared statistic).  

However, to obtain the spacecraft velocity as a function of time from this Doppler shift is not easy.  The codes must include all gravitational and time effects of general relativity to order $(v/c)^2$ and some effects to order $(v/c)^4$.  The ephemerides of the Sun, planets and their large moons as well as the lower mass multipole moments are included.  The positions of the receiving stations and the effects of the tides on the exact positions, the ionosphere, troposphere, and the solar plasma are included.  

Given the above tools, precise navigation was possible because, due to a serendipitous stroke of luck, the Pioneers were spin-stabilized. With spin-stabilization the craft are rotated at a rate of  $\sim$(4-7) rpm about the principal moment-of-inertia axis.  Thus, the craft is a gyroscope and attitude maneuvers are needed only when the motions of the Earth and the craft move the Earth from the antenna's line-of-sight.  Thus, especially in the later years, only a few orientation maneuvers were needed every year to keep the antenna pointed towards the Earth, and these could be easily modeled.  

Spin-stabilization is contrary to the cases of the later Voyagers,  which were 3-axis stabilized. 
With 3-axis stabilization, there are continuous, semi-autonomous, small gas jet thrusts to maintain the antenna facing the Earth.   This yields a navigation that is not as precise as that of the Pioneers.  

\begin{figure}[h!] 
    \begin{center}  
\includegraphics[width=3.75in]{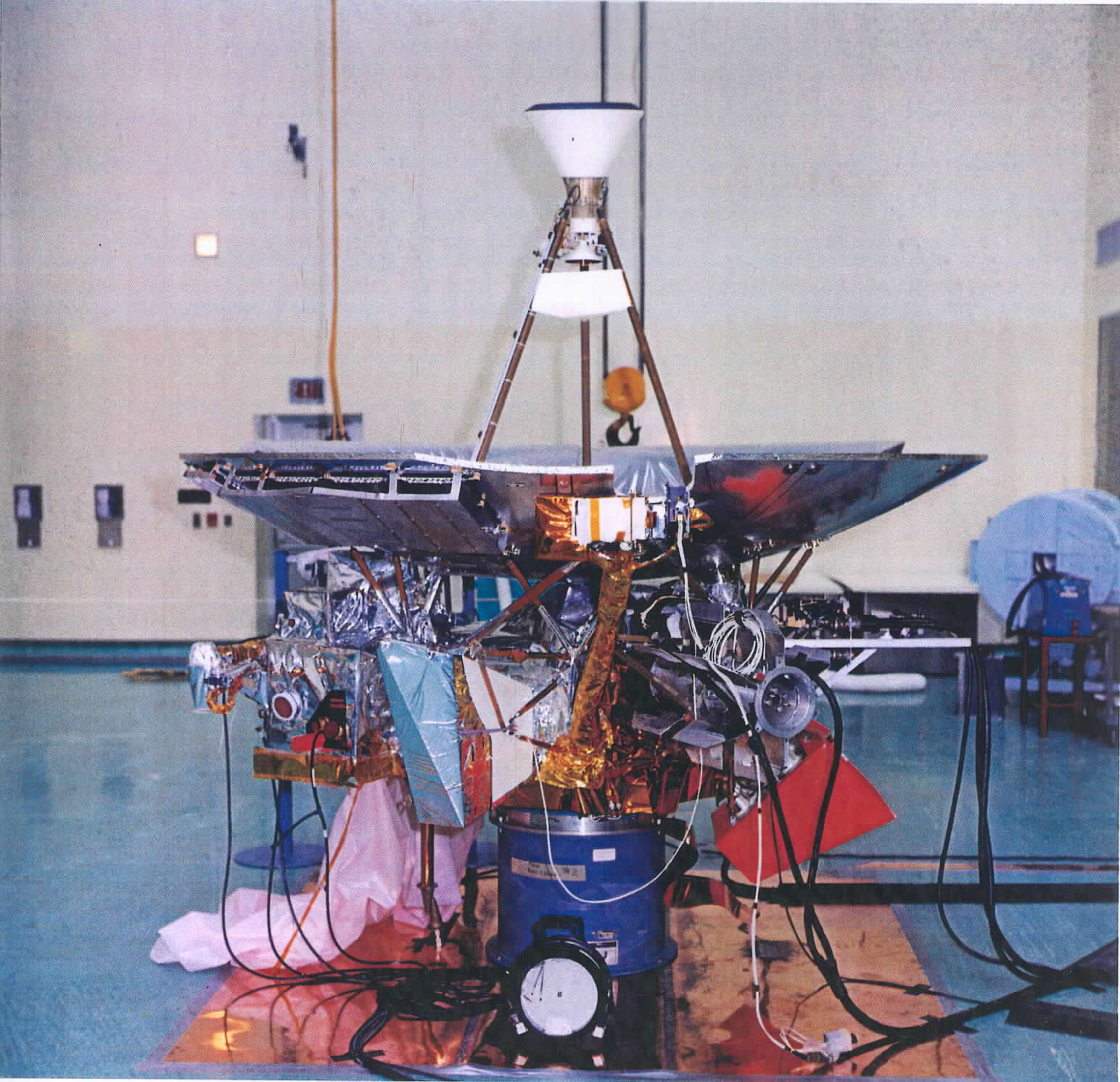}
\caption{NASA photo \#72HC94, with caption ``The Pioneer F spacecraft 
during a checkout with the launch vehicle third stage at Cape Kennedy.''
Pioneer F  became Pioneer 10 with the successful launch.
(The RTG booms are here withdrawn into launch mode.) 
\label{piopic}}
\end{center}
\end{figure}

The Pioneers were chosen to be spin-stabilized because of other engineering decisions.  As the craft would be so distant from the Sun solar power panels would not work.  Therefore these were the first deep spacecraft to use nuclear heat from $^{238}$Pu as a power source in Radioisotope Thermoelectric Generators  (RTGs).  Because of the then unknown effects of 
long-term radiation damage on spacecraft hardware (as well as a concern about contaminating the data from the scientific instruments), a choice was made to place the RTGs at the end of long booms.  This placed them away from the craft and thereby avoided most of the radiation that might be transferred to the spacecraft, especially the instrument bay. This was a fortuitous decision for purposes of doing precision celestial mechanics. 
The final fins on the RTGs were made larger than the original design, to increase the heat radiation away from the craft \cite{tele}.\footnote{
By contrast, radiation from the RTGs was not a concern on later spacecraft, such as the Voyagers and Cassini.  Thus, any science relying on trajectories approaching Pioneer accuracy is, for the Voyagers and Cassini, compromised by unmodeled thermal emission.    
} 
(Figure \ref{piopic} shows a photo of Pioneer 10 on the test stand.)  

Even so, there remained one relatively large effect on this scale that had to be modeled: the solar radiation pressure of the Sun, which also depends on the craft's orientation with respect to the Sun.  This effect is approximately 1/30,000 that of the Sun's gravity on the Pioneers and also decreases  as the inverse-square of the distance.  It produced an acceleration of  $\sim 20 \times 10^{-8}$ cm/s$^2$ on the Pioneer craft at the distance of Saturn (9.38 AU from the Sun at encounter).  (For comparison, the gravitational acceleration of the Sun at the Earth is 0.593 cm/s$^2$.)   Therefore,  any ``unmodeled force" on the craft could not be seen very well below this level at Jupiter.  However, beyond Jupiter it became possible.  

\newpage


\section{The Anomaly is Observed}
\label{discovery}

In 1969 one of us (JDA) became PI of the radioscience celestial mechanics experiment for the Pioneers.  He remained so until the official end of the extended mission \cite{extended} in 1997.   Working with Eunice Lau (who also later joined the Pioneer anomaly Collaboration),  the Pioneer Doppler data going back to 1976 for Pioneer 11 and 1981 for Pioneer 10 (but also including the Jupiter flyby) was archived at the  National Space Science Data Center (NSSDC), something that later was to prove extremely helpful.  

Part of the celestial mechanics effort, working together with the navigation team, was to model the trajectory of the spacecraft very precisely and to determine if there were any unmodeled effects.   

\begin{figure}[h!] 
    \noindent
    \begin{center}  
\includegraphics[width=3.75in]{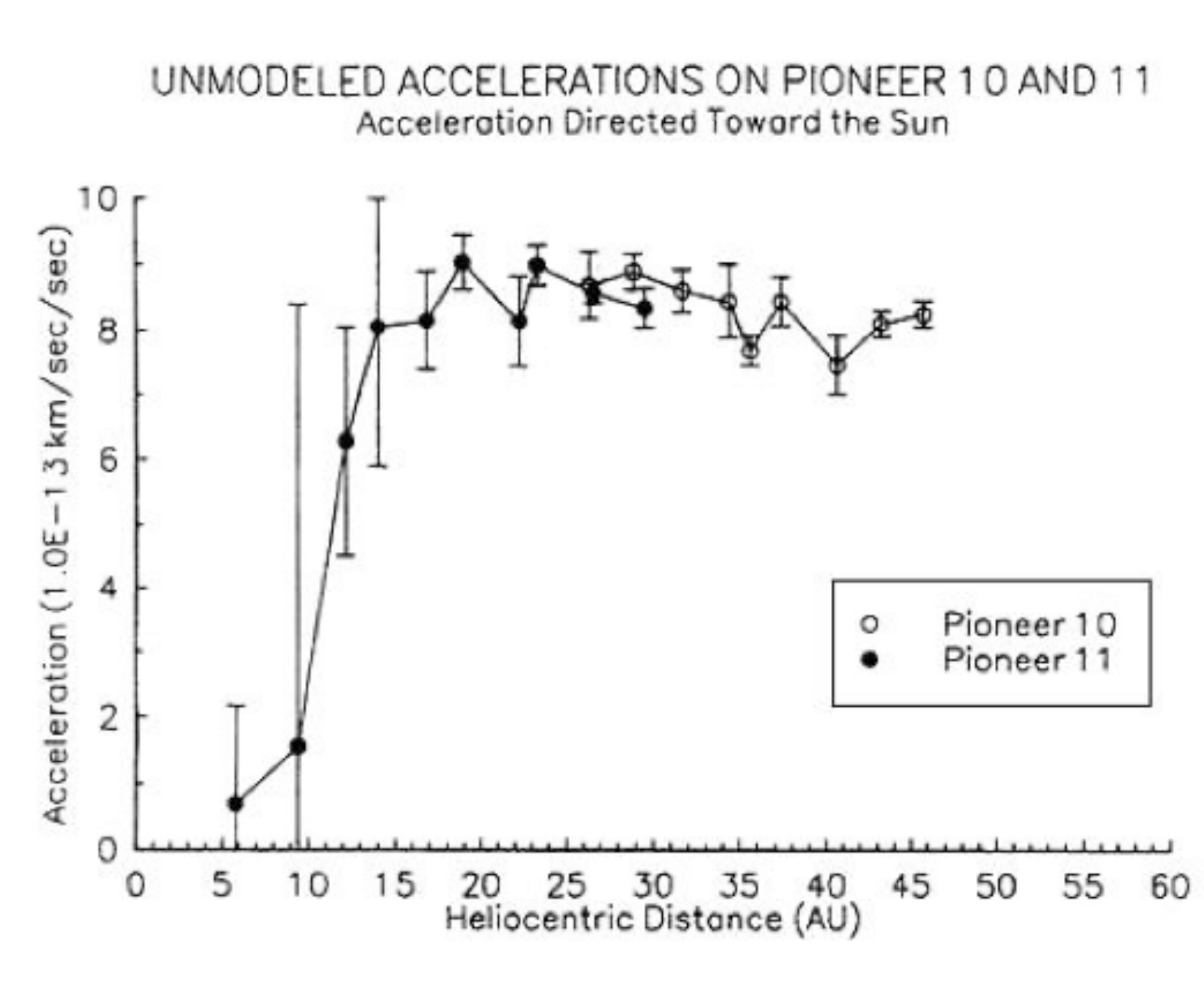}
\caption{A JPL Orbital Data Program (ODP) plot of the early unmodeled accelerations of Pioneer 10 and Pioneer 11, from about 1981 to 1989 and 1977 to 1989, respectively.   This graph first appeared in JPL memos from the period 1992 \cite{JPLmemo}.  
\label{correlate}}
\end{center}
\end{figure}

After 1976 small time-samples (approximately 6-month to 1-year averages) of the data were periodically analyzed.  (This was especially true for Pioneer 11 which was then on its Jupiter-Saturn cruise.)  
These data points were obtained individually by a number of very-qualified investigators,\footnote{This data taking was possible because, after their successes, an extended mission was decided upon for the Pioneers \cite{extended}.
}
including J. D. Anderson, J. Ellis,  E. L. Lau, N. A. Mottinger, G. W. Null, and S. K. Wong.
At first nothing significant was found \cite{null76}.    
But when a similar analysis was done around Pioneer 11 's Saturn flyby, things dramatically changed.  (See the first three data points in Fig. 
\ref{correlate}.)  So people kept following Pioneer 11.  They also started looking more closely at the incoming Pioneer 10 data.  


By 1987 it was clear that an anomalous acceleration appeared to be acting on the craft with a magnitude $\sim 8 \times 10^{-8}$ cm/s$^2$, directed approximately towards the Sun    The effect was a concern, but the effect was small in the scheme of things and did not affect the necessary precision of the navigation.  But by 1992 it was observed that a more detailed look would be useful \cite{JPLmemo}.\footnote{This reference contains the original 
Figure \ref{correlate}.}

{\bf (*** LOCATION OF SIDE BAR II. ***)}


\section*{SIDE BAR II:  Pioneer 11 at Saturn}

{\it
It is technically imprecise to consider the total energy of the Pioneer 11 spacecraft separately from that of the rest of the solar system.   Nonetheless, consider 
the 4-body problem with i) the solar-system barycenter (differing from the center especially because of Jupiter) as the origin of inertial coordinates and with ii) 
the potential energy given by the Sun, Saturn including its leading multipoles (up to octapole), and Titan. 
One can then ask \cite{hale}, ``What is the total energy of the spacecraft, $E_P$, with time?"    

While at Saturn, the time histories for Pioneer 11's kinetic energy, Saturn's contribution to the potential energy, the total energy, and the determined value of the Jacobi ``constant" (all per unit mass) are given in 
Figure \ref{11totE}. 
(The potential from the Sun varied much more slowly and basically was a bias of $-94.5$ (km/s)$^2$.)  


\begin{figure}[h!] 
    \noindent
    \begin{center}  
\includegraphics[width=2.5in]{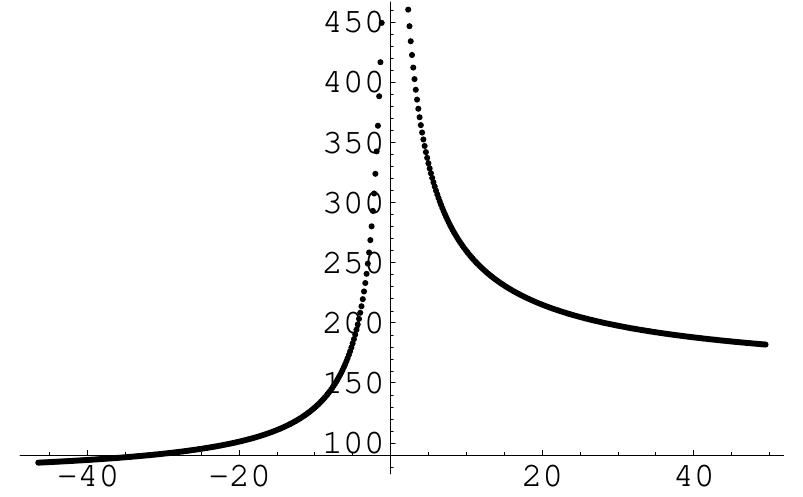} 
\includegraphics[width=2.5in]{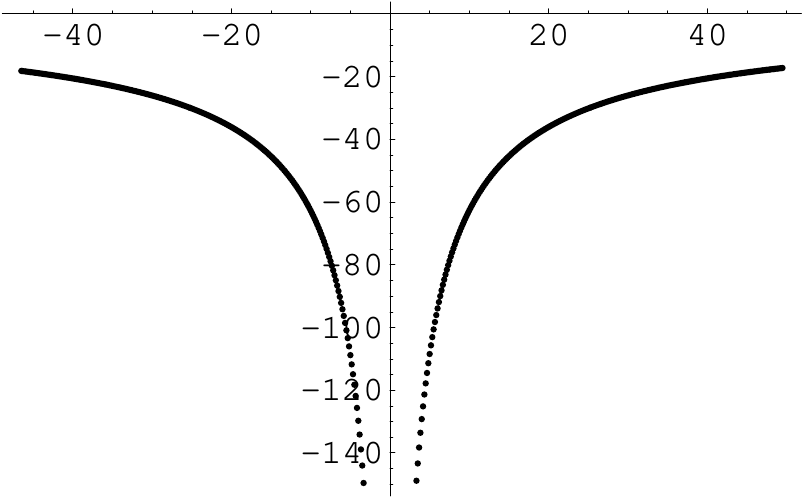}  \\
\vspace{.25in}
\includegraphics[width=2.5in]{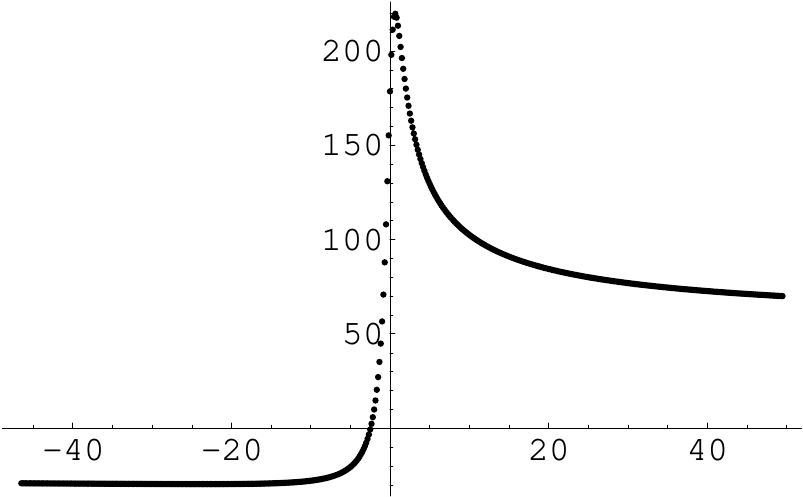} 
\includegraphics[width=2.5in]{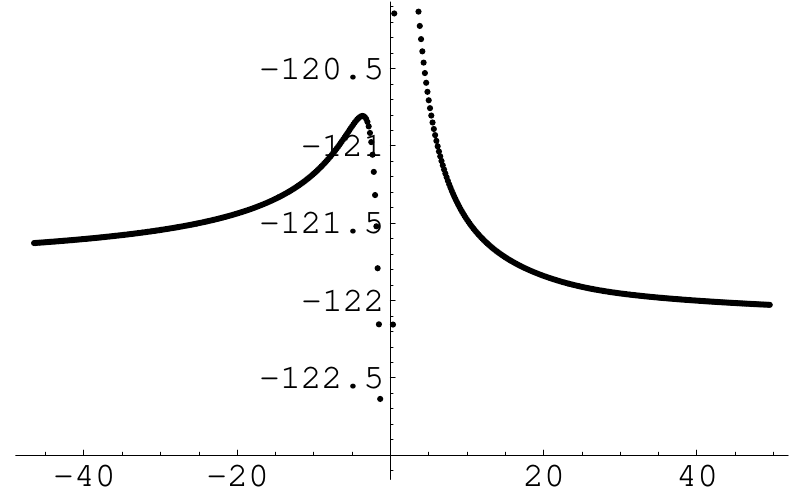} 
\caption{Energies per unit mass of the Pioneer 11 spacecraft in (km/s)$^2$ plotted vs. time about closest approach to Saturn in hours: {\it Upper Left}: the kinetic energy (${\cal K}_P = K_P/m_P$); {\it Upper Right}:  Saturn's contribution to the potential energy; {\it Lower Left}: the total energy (${\cal E}_P = E_P/m_P$): and {\it Lower Right}: the determined value of the Jacobi ``constant" ($J$) computed using Eq. (\ref{jacobi}).  The latter shows the small residuals about the fit. 
\label{11totE}}
\end{center}
\end{figure}


Figure \ref{11totE}   
might surprise even though, as we come to below, it was to be expected.  Although in the Saturn-centered frame one expects everything to be symmetric before and after encounter, going to the solar system barycenter frame, to first approximation one {\it a priori} might anticipate a continuous, monotonic transfer of energy with time.  There would be a steady monotonic increase of energy before encounter and then a smooth still-monotonically increasing transition to a new constant energy after encounter. That is not seen.

Figure \ref{11totE}'s up-down asymmetry is huge and there is nothing like left-right symmetry.  Pioneer 11 reached a state of positive total energy about 2-1/2 hours before closest approach to Saturn. But then the spacecraft first gained more than the final energy-shift from the planet and then lost some.  Does this make sense?  If we go back to our previous aside it does.

Reconsider 
Figure \ref{flandrofig} 
and Eqs. (\ref{fly1}) and (\ref{fly2}).  These last are
asymptotic formulae. But we can make them time-dependent formulae by changing ${\bf v}'_f$ to ${\bf v}'(t)$.  Then, we have 
\begin{equation}
\Delta {\cal K}(t) = {\bf v}_p \cdot ({\bf v}'(t) - {\bf v}_i').
\end{equation}
The second term is a constant.    By looking at 
Figure \ref{11totE}  
one can see that the first graph shows the spike in after perigee with respect to the planet being flown by.  
As the satellite goes around the back side of the planet, 
${\bf v}'(t)$ starts to align with ${\bf v}_p$.  The maximum  of $\Delta {\cal K}$ is reached when ${\bf v}'(t)$ is parallel with ${\bf v}_p$.  This occurs just after perigee.  Then as the satellite swings around further, ${\bf v}'(t)$ goes out of alignment with ${\bf v}_p$ so $\Delta {\cal K}(t)$ decreases some.  

The first three graphs of 
Figure \ref{11totE} 
now makes sense. (In particular, ${\cal E}(t)$ is calculated as a function of time and is shown in the third graph of Figure \ref{11totE}.)  
The last graph of Figure \ref{11totE}   shows that the energy comes from angular momentum (${\cal L}$).   (See Eq. \ref{jacobi}.)  
It is clear what would happen if the orbit went the other way (gravity-decrease of total energy).
}\\


For independent reasons, in 1994 MMN contacted Anderson to ask about how well we understood gravity far out in the solar system.  Nieto was specifically thinking of the Voyagers, but for reasons explained elsewhere in this article, Anderson went to the Pioneers.  When MMN read JDA's email statement that,``By the way, the biggest systematic in our acceleration residuals is a bias of $8 \times 10^{-13}$ km/s$^2$ directed toward the Sun," he almost fell off of his chair.\footnote{MMN was on a wheeled chair in an office with a tile floor.  So upon reading Anderson's statement he spontaneously arched his back exclaiming, ``What?" thereby almost causing a calamity.}

The result was an announcement in a 1994 Conference Proceedings \cite{bled}. The strongest immediate reaction was that the anomaly could well be an artifact of JPL's Orbital Determination Program (ODP), and could not be taken seriously until an independent code had tested it.  So Anderson put together a team that included his collaborator, Eunice Lau, as well as two former Pioneer co-workers Phil Laing and Tony Liu (see Figure \ref{phil}), 
who were then associated with The Aerospace Corporation. The final member of the team was Slava Turyshev, a postdoc whom we both agreed should be welcomed.  Laing and Liu used the independent CHASMP navigation code they had developed to look at the Pioneer data.  To within small uncertainties, their result was the same as that obtained by JPL's ODP.   

The Pioneer anomaly Collaboration's discovery paper appeared in 1998 \cite{pioprl} and a detailed analysis appeared in 2002 \cite{pioprd}. The latter used Pioneer 10 data spanning 3 January 1987 to 22 July 1998 (when the craft was 40 AU to 70.5 AU from the Sun) and Pioneer  11 data spanning 5 January 1987 to 1 October 1990 (when Pioneer 11 was 22.4 to 31.7 AU from the Sun).  This analysis was the first to look at all the systematics issues in detail \cite{pioprd,piompla,pioajp}.  The largest systematics are shown in Table \ref{error_budget}.


\begin{table*}[ht]
\begin{center}
\caption{Error Budget: A Summary of the Largest Biases and Uncertainties.
\label{error_budget}} \vskip 20pt
\begin{tabular}{rlll} \hline\hline
Item   & Description of error budget constituents & 
Bias~~~~~& Uncertainty    \\     &       &
           $10^{-8} ~\rm cm/s^2$ &  $10^{-8} ~\rm cm/s^2$  \\\hline
   &                 &             &         \\
 1 & {\sf Systematics generated external  to the spacecraft:}  && \\
   & a) Solar radiation pressure    &  $+0.03$   & $\pm 0.01$\\
   & b) Solar corona      & & $ \pm 0.02$ \\ [10pt]
 2 & {\sf On-board generated systematics:} &&         \\
   & a) Radio beam reaction force         & $+1.10$&$\pm 0.10$ \\
   & b) RTG heat reflected off the craft      &  $-0.55$&$\pm 0.55$ \\
   & c) Differential emissivity of the RTGs         & &  $\pm 0.85$ \\
   & d) Non-isotropic radiative cooling of the spacecraft && $\pm 0.48$\\
   & e) Expelled Helium  produced  within the RTGs 
           &$+0.15$ & $\pm 0.16$   \\
   & f) Gas leakage           &         &  $\pm 0.56$   \\
   & g) Variation between spacecraft determinations  
           &   $+0.17$  & $\pm 0.17$  \\[10pt]
 3 & {\sf Computational systematics:}          &&         \\
   & a) Accuracy of consistency/model tests          & &$\pm0.13$ \\
  & c) Other important systematics      & &$\pm 0.04$ \\[10pt]  
\hline
   &                              &&         \\
   & Estimate of total bias/error     & $+0.90$& $\pm 1.23$       \\
   &                              &&         \\
\hline\hline
\end{tabular} 
\end{center}
\vskip -10pt 
\end{table*}


The error is mainly systematic (the statistical error is minimal). . 
The most significant biases and errors (all in units of $10^{-8}$ cm/s$^2$) are \cite{pioprd,piompla,pioajp}:
radio beam reaction force $(1.10\pm 0.11)$; RTG heat reflected off the craft $(-0.55\pm 0.55)$; differential emissivity of the RTGs $(\pm 0.85)$; non-isotropic radiative cooling of the craft $(\pm 0.48)$; gas leakage $(\pm 0.56)$.    
Note that except for the radio beam reaction force, which mainly is a bias, and the gas leak uncertainty, which mainly is evidence of shifts back and forth in the constancy of the anomaly after certain leak events \cite{pioprd}, the other major systematics are due to heat in some form or the other.  

When added to the pure experimental residuals, which would indicate an acceleration of  $a_{\tt expt} = (7.84 \pm 0.01) \times 10^{-8}$ cm/s$^2$, this yields the final result for the anomaly, that there is an unmodeled acceleration, directed approximately towards the Sun, of 
\be
a_P = (8.74 \pm 1.33) \times 10^{-8} \mathrm{cm/s}^2.
\ee

Two later and independent analyes were done of this data \cite{craig,oystein}.  They obtained similar results.  The conclusion, then, is that this ``Pioneer anomaly" is in the data.  
The question is, ``What is its origin?"



\section{Proposed Origins of the Anomaly}
\label{meaning}


\subsection{On board systematics}
\label{syst}

The analysis of the Pioneer Doppler data \cite{pioprl,pioprd} used  modern data from 1987.0 to 1998.5 for Pioneer 10 and from 1987.0 to 1991.0 (when coherent data was lost) for Pioneer 11.  This was done for three main reasons:  i) The modern data was still easily accessible and it was stored on current platforms that used existing software.  (In contrast, older data where available was in increasingly obsolete formats the further back in time one went.)  ii) During the time of the modern data acquisition the crafts were farther away from the Sun (greater than 40 and 20 AU, respectively, for Pioneers 10 and 11).  Thus, solar radiation pressure and plasma effects were a smaller complicating factor.  iii) Further out from the Sun, there were fewer Earth-reorientation maneuvers needed to re-point the high-gain antenna towards the Earth.  Any maneuvers need to be modeled in the orbit determination.   To the accuracy of the analysis, the anomaly was constant, but this accuracy was only $\sim$15\%.  

It is tempting to assume that radiant heat must be the cause of the acceleration, since only 63 W of directed power could cause the effect (and much more heat than that is available).  This 63 W can be understood from the simple formula
\be 
a_{\tt Rh} = \frac{P}{MC}
\ee 
where $P$ is the directed power, $M$ is the mass of the craft, and $c$ is the velocity of light.  

The heat on the craft ultimately comes from the  Radioisotope Thermoelectric Generators (RTGs), which yield heat from the radioactive decay of $^{238}$Pu.  Before launch,
the four RTGs had a total thermal fuel inventory of 2580 W ($\approx
2070$ W in 2002). 
Of this heat 165 W was converted at launch into electrical power emanating from around the main bus ($\approx 65$ W in 2002), significantly from the louvers at the bottom of the 
bus.\footnote{
The 30 louvers on each craft are highly reflective shutters.  They are attached to the craft by temperature activated bimetallic springs which are thermally coupled radiatively to the craft platform.  These are designed to be completely open (and thereby let the most heat out of the equipment bays) above $\sim 90^\circ$ F and to be completely closed (and thereby keep in the most heat) below $\sim 40^\circ$ F.
}

Therefore, it has been argued \cite{katz,uskatz} whether the anomalous
acceleration is due to anisotropic heat reflection off of the back of the
spacecraft high-gain antennas, 
It has been further argued \cite{murphy,usmurphy} whether the heat emanating from the louvers (open or closed) is the origin of the anomaly, and if a combination of both these sources must be considered \cite{piompla,scheffer}. 

In fact, the craft was designed, again serendipitously, so that the heat was radiated out in a very fore/aft symmetric manner.   Further, the heat from electric power went down by almost a factor of 3 during the mission.  
With all these points in mind, and even though admittedly heat is the most likely explanation of the anomaly, no one as yet has been able to firmly tie this down, despite the heated controversy we have referred to \cite{piompla},
\cite{katz}-\cite{scheffer}. 
Therefore, heat as a mechanism yielding an approximately constant effect remains to be clearly resolved, but studies are underway (see Section 
\ref{edata}).

Indeed, from the beginning we observed that a most likely origin is directed heat radiation \cite{pioprl,pioprd}.  However, suspecting this likelyhood is different from proving it.  Even so, investigation may well ultimately show that heat was a larger effect than originally demonstrated.


\subsection{Other physics}

Drag from normal matter dust \cite{drag} as well as gravity from the Kuiper belt have been ruled out \cite{pioprd,boss,kuiper,bkuiper}.  Also, if this is a modification of gravity, it is not universal; i.e., it is not a scale independent force that affects planetary bodies in bound orbits \cite{pioprd,rathke,iorio}.
The anomaly could, in principle be i) some modification of gravity \cite{serge}-\cite{moffat}, ii) drag from dark matter \cite{drag,volkas} or a modification of inertia \cite{mond}, or iii) a light acceleration \cite{asubt}.  (Remember, the signal is a Doppler shift which is only interpreted as an acceleration.)  Future understanding of the anomaly will determine which, if any, of these proposals are viable.

In the above circumstances the true direction of the anomaly should be \cite{origin} i) towards the Sun, ii) along the craft velocity vector,\footnote{
Technically it is along the spacecraft's change in momentum.
} 
or iii) towards the Earth.  If the origin is heat, or any other spacecraft-generated force, the acceleration would be iv) along the spin axis. (Any internal systematic forces normal to the spin axis are canceled out by the rotation.)


\section*{SIDE BAR III:  Other Anomalous Dynamical Effects}
\label{sideIII}

{\it
Looking carefully at Figure \ref{correlate}, one can note a small apparent annual oscillation on top of the constancy.  Careful analysis of clean, late-time data showed \cite{pioprd} that this signal is indeed significant, {\it as well as} a diurnal signal.  These are believed to be independent of the main anomaly and perhaps due to improper orientation of the direction from the Earth.  However, the question is not settled yet.  

Interestingly, in addition to Pioneer 10 at Saturn, there is another anomaly that is associated with planetary flybys, specifically Earth flybys.   
During such a flyby the
total energy and angular momentum of the solar system are
conserved.  Further, independent of the heliocentric energy change of the craft itself, 
the spacecraft's total {\it geocentric} orbital
energy per unit mass {\it should} be the same before and after the flyby.  
But even after perturbations are accounted for, the data indicates this is not always true.  

Instead, for at least the Earth flybys by the Galileo (GEGA-I) \cite{gega1}, NEAR \cite{ant}, and Rosetta \cite{trevor} spacecraft,
the geocentric orbital energies after the closest approach to Earth
were noticeably greater than the orbital energies before closest
approach.  Further, the changes were much too large for any conventional time-explicit cause, specifically from the Earth's longitudinal harmonics or the motions of the Moon and Sun.  In particular, for hyperbolic excess velocities with respect to the Earth of $v_\infty =$(8.949, 6.851 3.863) km/s, respectively, the unmodeled changes (taken to occur at periapsis) were $\Delta v_{flyby}=(3.02\pm 0.08,~13.46\pm 0.13,~ 1.82\pm 0.05)$ mm/s 
So far, no mechanism, either external or internal
to the spacecraft, that could produce these observed net changes in
orbital energy has been identified \cite{hale,claus}.

This is discussed in detail elsewhere, including with energy transfer diagrams \cite{hale}.  There is hope that data from the Rosetta flyby of Mars on 25 February 2007 could provide further information.
}


\section{Short Term Possibilities for Progress}
\label{searching}


\subsection{Studying the entire data set}  
\label{edata}

As explained in Sec. \ref{syst}, the major analysis \cite{pioprd} used data from 1987.0 to 1998.5.  However, to well discern whether the anomaly is truly constant or if there is at least a piece with the time-dependence characteristic  of the either the 87.74 year half-life of $^{238}$Pu and/or the more rapid decay of the electrical power, a long data span would be of help \cite{edata}.  Further, if the effects of solar radiation pressure, solar plasma, and many maneuvers that occur close in to the Sun could be disentangled, one might be able to do 3-dimensional tracking precisely enough to determine the exact direction of the anomaly.  Perhaps most intriguingly, by closely studying the data around Pioneer 11's Saturn flyby (and Pioneer 10's Jupiter flyby) it could be determined if, indeed, there was an onset of the anomaly near these transitions to hyperbolic-orbits.  

If all the Doppler data, from launch to last contact, could be precisely analyzed together {\it and} all systematics external to the craft could be separated, the above tasks could be accommodated.  Indeed, one of the main goals of current work is to analyze as much of the Pioneer data as possible to look more precisely for time-dependent changes and the true direction of the anomaly.  However, this is much easier said than done.   The data itself was stored using obsolete formats on obsolete platforms.  Further, it was not to be found in one place, although the NSSDC did have a sizable segment.  The untangling and interpretation of the archival Doppler data is an ongoing and important project \cite{TT,TT2}. 

Further, there is the telemetry.  Over the years, Larry Kellogg of Ames had the foresight and dedication to compile and retain obsolete-format telemetry files containing the engineering data, including such things as temperatures, voltages, spin rates, etc.   This data has now been transferred to modern format  \cite{TT,TT2}.  

In the long run the study of the telemetry might prove to be of most ``use."  The Collaboration has long observed that, even if the anomaly turns out to be due to systematics, a thorough anomaly inquiry would still result in a win.  One  would obtain a better understanding of how spacecraft behave in deep space and therefore how to build, model, and track craft there.  It is a case of reverse engineering.  

Specifically, can one model the heat transfer in a craft well enough to know before hand what the effects will be?  This task has proved notably difficult in the past, with predictions deviating markedly from observations \cite{cassini}.  However, techniques are improving and convincing results could be obtained.   

If fact, return again to Figure \ref{correlate}.  
If the first data point is correct, and not due to misrepresentation of radiation pressure and maneuvers, it could indicate a  small systematic of order $2 \times 10^{-8}$ cm/s$^2$.  Then the second point would represent an onset and the third the full flowering of the anomaly.  (The most frustrating conclusion would be half systematics and half a signal of something else.)  

Thus, although the telemetry will aid in the final detailed conclusions, ultimately things will be decided by the analysis of the Doppler data, showing both the size of the anomaly with time and also, with luck, its direction.  This latter will be most difficult to discern because of the large systematics close in to the Sun.    
The analysis will be further complicated by problems from using old archival and sometimes forgotten formating standards.  But the results could be very rewarding.


\subsection{The New Horizons mission to Pluto}
\label{newhorizons}

On 19 Jan 2006 the New Horizons mission to Pluto and the Kuiper Belt was launched from Cape Canaveral.  Although of relatively low mass ($\sim$478 kg including hydrazine thruster fuel) and with high velocity ($\sim$20-25 km/s), this craft was not designed for precision tracking.  Even so, serendipitously it might be able to yield useful information \cite{newH}.

The main reason is that for much of its life New Horizons will be in spin-stabilization mode, which is good for tracking.  It was in spin-stabilization mode for part of the six months before the Jupiter observing period (January-June, 2007, with encounter on 28 Feb. 2007).  It also will be spin-stabilized for much of the period after June 2007 until soon before the Pluto encounter on  14 July 2015.  This is designed to save fuel so that as much fuel as possible will be available after Pluto encounter to aim at a Kuiper Belt Object.  The Doppler and range data from these periods could supply a test, at some level, of the Pioneer anomaly, especially since the velocity of the craft before ($\sim$21 km/s) and after ($\sim$25 km/s) the Jupiter encounter will be significantly different that those of the Pioneers ($\sim$11 km/s).  With luck something could be learned from the New Horizons data by 2010 or soon thereafter.    
   
However, it is not clear that the data will be precise enough for our purposes.
The first problem will be the on-board heat systematics.  The large RTG is mounted on the side of the craft, and produced $\sim$4,500 W of heat at launch.  A rough calculation shows that a systematic of $\sim 20 \times 10^{-8}$ cm/s$^2$ or larger will be produced.  Since, as we discussed above, the modeling of heat systematics is  difficult, understanding this systematic would be an important problem to overcome. 

However,  soon after launch a 180 degree “Earth acquisition maneuver” rotation was performed.  (The reason for this maneuver  was to aim the main high-gain antenna towards the Earth.).  The difference in the Doppler shift immediately before and after the rotation could in principle yield a difference measurement of the heat acceleration.  This is because the fore-aft heat acceleration would be pointed first in one direction and then in the opposite.  Unfortunately, a good  determination may be difficult because of the high solar radiation pressure (which will vary somewhat in the two orientations), the relatively small data set before the maneuver, and finally because of the difficulties in modeling the acceleration caused by the maneuver and any attendant gas leakages.  

In summary, we await to see if this mission can be of use to the study of the anomaly.  The analysis of data from the Pluto mission will be a challenge.   Even though it might be nonproductive for our purposes, we encourage it anyway.


\section{A Dedicated Mission?}
\label{esa}

If the above efforts are not able to yield a {\it negative} resolution of the anomaly, then a new experimental test might be needed, either as an attached experiment or probe, or even as a dedicated mission \cite{rathke,origin}.  An example of the former would be a module first attached to a high-velocity mission to very deep space, such as the InterStellar Probe concept.  Further out than at least the orbit of Jupiter or Saturn, the module would be detached and fly on autonomously.  The module would then be tracked from the ground.

As discussion of the anomaly and a possible mission was growing \cite{origin}, there simultaneously arose an interest in the problem in Europe.  In May 2004 a meeting was held at the University of Bremen to discuss the anomaly, and from this an international collaboration was formed to propose a dedicated test of the anomaly \cite{cosmic}.   Institutions from all over Europe, including from France, Germany, Great Britain, Italy, Netherlands, Portugal, and Spain, have joined the collaboration.  

The goal is to present a proposal as a Theme for ESA's Cosmic Vision program, with launches to occur during the period 2015-2025.  As such its timing would be perfect if the two investigations described above indicate that a dedicated test of the anomaly is called for.  A driving consideration would be new technology.   The mission would take insight from knowledge of what allowed the Pioneer craft to be navigated so well and add to it.   

The collaboration first prepared and published a preliminary proposal \cite{cosmic}.  
The aim was to accurately determine the heliocentric motion of a test-mass utilizing 2-step tracking with common-mode noise rejection.  
A state-of-the-art Ka-band ($\sim 32-35$ GHz at the DSN) tracking system, using both Doppler and range, was to be used to track the main satellite to an accuracy approaching $0.1 \times 10^{-8}$ cm/s$^2$.  
This main craft would be spin-stabilized.  This would eliminate the need for many maneuvers to maintain antenna orientation and, because of the rotation, any systematic accelerations not in the fore/aft direction would be averaged out to zero.  To reduce heat acceleration of the craft, the design would be extremely fore/aft symmetric as far as heat/power-radiation were concerned.  (The heat would be radiated out in fore/aft and  axially symmetric manners, a symmetry proposed earlier \cite{origin}.)   

\begin{figure}[h!] 
    \begin{center}  
\includegraphics[width=3.75in]{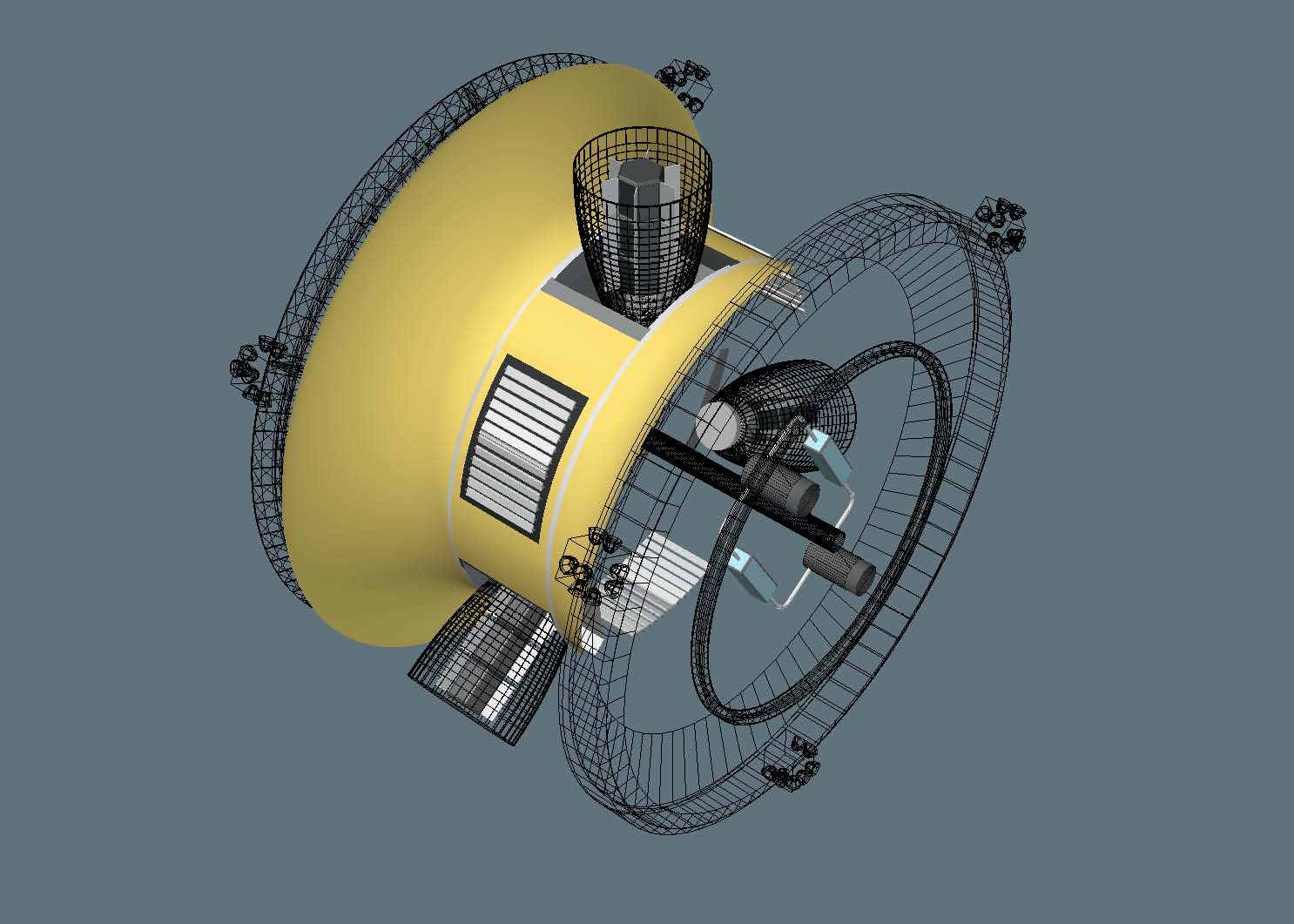}
\caption{A schematic, cut-away drawing of the preliminary ``Pioneer Anomaly Explorer" concept.  The side facing away contains the radio antenna to communicate with Earth.  On the facing side are the mW laser and the canisters that will emit the spheres covered with corner-cubes.  (Drawing courtesy of Alexandre D. Szames.) 
\label{ESAcraft}
}
\end{center}
\end{figure}

In Figure \ref{ESAcraft}
we show a schematic cut-away of this preliminary model.  The side exterior surfaces are curved to symmetrically reflect and radiate heat from the side of the bus.  (Heat from inside the equipment bus would come out of louvers located between the RTGs).  This also would  symmetrically reflect heat from the RTGs which are in parabolic Winston cone reflectors.   The RTGs could be extended out on booms after launch.


The second stage in the tracking was to take place after   
the main craft was in deep space.  This would be when no further trajectory modification acceleration maneuvers would occur, be they from a final stage chemical rocket, a planetary flyby, thrusters, or even a to-be-jettisoned solar sail. 
This would probably be at a distance of 5-10 AU, when the craft hopefully had a velocity of $>$5 AU/yr.  From then on, and especially at distances of  25-45 AU, when solar radiation pressure is reduced, precise data could be taken.

Then a ``formation flying" system would be used.  From the forward side facing away from the Earth small spheres, covered with corner-cube reflectors, would be sent out to be tracked from the main satellite with mW laser ranging.  The passive spheres would be at a distance of order $>$500 m from the main satellite, which satellite would utilize occasional maneuvers to maintain formation.   This final step could yield an acceleration precision approaching $10^{-10}$ cm/s$^2$.  On board one could also carry sensitive drag-free DC accelerometers.  
In Figure \ref{ESAcraft} one can also see where the spheres would be extruded and the central location of the laser.  

But after much deliberation, the collaboration has tentatively decided instead to submit what amounts to a two-stage proposal.  The first will be a more modest mission along the lines of the EADS Astrium Enigma small-mass concept \cite{enigma}.  It will use modern 3-axis, drag-free accelerometers  developed by ONERA in France \cite{onera}.  This mission would be a slower mission using flybys for gravity assists out to $\sim 20$ AU.  

This mission would get its main power from solar cells.  (As aids, "small RTGs" or radioisotope heater units might be added.)   As such this mission could be flown cheaper and quicker.  But it would have power limitations, entail large solar radiation pressure effects from the solar-cell assemblies needed at larger heliocentric distances, and it would not achieve hyperbolic orbit or large heliocentric distances.

We note that large solar arrays on spacecraft flying by the Earth could well make a precise test of the flyby anomaly very difficult. It would be preferable to fly the entire mission on RTGs.  This statement is not a stance on the politics of nuclear proliferation as the $^{238}$Pu used in RTGs is {\it not} the isotope desired from reactors to be used in weapons. This isotope is simply a heat source for converting heat energy into electric energy for the spacecraft.  

A later mission would be equipped with modern quantum-technology instruments to measure precise accelerations at longer distances.  
If called for, dedicated missions such as these could be very exciting and definitive.  Studies of them should proceed as, during the next few years, a determination is made of their requirement.


\section{Conclusion}
\label{conclusion}

That the Pioneer anomaly is a physical effect is no longer in doubt.  The only question is its origin. What we have here is perhaps analogous to other questions involving not-understood possible gravitational effects, from early tests of Newton's law to current discusions over ``dark matter" vs a modification of gravity \cite{mond}.

Possibly the most famous dealt with the discovery of Neptune.  By the 1840's it was clear that the orbit of Uranus was not understood.  One of the greatest triumphs in celestial mechanics history was the independent solutions of John Couch Adams and Urbain Jean Joseph le Verrier of the inverse problem, yielding successful predictions of the location of the quickly discovered Neptune \cite{neptune}.\footnote{
An input into the solution \cite{neptune} was what amounted to the Titius-Bode Law of Planetary Distances \cite{n1,n2}.  
}  
This was a triumph of ``dark matter" over a modification of gravity.  

Thereupon le Verrier started a complete study of all the planets.  When he returned to Mercury in 1859, he again found an earlier troubling problem \cite{leV2}, the precession of Mercury's perihelion was too large, by 33-38 arc-seconds/century.  (Later Simon Newcomb did a more precise calculation and found the ``modern" value of 43.)

Once again the question was the origin of the anomaly. 
le Verrier was convinced of the absolute correctness of Newton's theory, and therefore went for the ``dark matter" solution: 
the introduction of an undiscovered planet Vulcan orbiting close to the Sun \cite{vulcan}. 
He remained convinced of this to his death, despite continued lack of observational success.   This time the resolution was a modification of gravity, to general relativity. In the first success of general relativity, Einstein found that it yielded Newcomb's result. 

What about the Pioneer anomaly?  
It is becoming increasingly clear that not only does the Pioneer anomaly not affect planets in the inner solar system, it also does not appear to affect planets in the outer solar system and most likely not Oort cloud comets or recently discovered Kuiper Belt Objects (KBOs) either. It will be difficult for a simple modification of the Newtonian inverse-square law or a new metric theory of gravity, both of which demand a universality of free fall, to explain the Pioneer anomaly. If one invokes a new metric theory (as an extension of the Einstein idea) and thereby fits the Pioneer anomaly, that attempt will very likely play havoc with solar-system dynamics somewhere else; for example an obvious failure to fit observations of natural bodies in the outer solar system. If the Pioneer anomaly is new physics, something more subtle is most likely involved. 

We therefore have to conclude that it is unlikely that the Pioneer anomaly is caused by new physics. Such discoveries are rare. However, it is not ruled out.  It is possible that the Pioneer anomaly could be something importantly new.  For that reason, and also because we want to be able to account for every source of systematic error in deep space navigation, we continue to pursue the study of spacecraft trajectories, whether in deep space or during planetary encounters. We recommend that you stay tuned. This anomaly will be resolved eventually, one way or the other.\footnote{The work described in this manuscript was supported by the US Department of Energy (MMN) and by the National Aeronautics and Space Administration (JDA).}



\end{document}